\RequirePackage{fix-cm}
\documentclass[smallextended]{svjour3}       
\usepackage{graphicx}
\usepackage{url}
\usepackage{subfigure}
\usepackage{array}
\usepackage{fp}
\usepackage{verbatim}
\usepackage[titletoc]{appendix}
\usepackage{multirow}
\usepackage{makecell}
\usepackage{epstopdf}
\usepackage{color}
\usepackage{graphics}
\usepackage{epsfig}
\usepackage{latexsym}
\usepackage{ifthen}
\usepackage{fancyhdr}
\usepackage{verbatim}
\usepackage{amsfonts}
\usepackage{amsmath}
\usepackage[round]{natbib}
\newcommand{\thickhline}{\noalign{\hrule height 1pt}}

\begin{document}

\title{Query Segmentation for Relevance Ranking in Web Search
}


\author{Haocheng Wu         \and
        Yunhua Hu \and
        Hang Li \and
        Enhong Chen
}


\institute{H. Wu \at
              University of Science and Technology of China \\
              \email{ustcwhc@outlook.com}           
           \and
           Y. Hu \at
              Alibaba.com   \\
              \email{wugou.hyh@taobao.com}
           \and
           H. Li \at
                Noah's Ark Lab of Huawei Technologies\\
              \email{hangli65@gmail.com}
           \and
           E. Chen \at
            University of Science and Technology of China \\
              \email{cheneh@ustc.edu.cn}           
}

\date{Received: date / Accepted: date}

\maketitle

\begin{abstract}
In this paper, we try to answer the question of how to improve the state-of-the-art methods for relevance ranking in web search by query segmentation. Here, by query segmentation it is meant to segment the input query into segments, typically natural language phrases, so that the performance of relevance ranking in search is increased. We propose employing the re-ranking approach in query segmentation, which first employs a generative model to create top $k$ candidates and then employs a discriminative model to re-rank the candidates to obtain the final segmentation result. The method has been widely utilized for structure prediction in natural language processing, but has not been applied to query segmentation, as far as we know. Furthermore, we propose a new method for using the result of query segmentation in relevance ranking, which takes both the original query words and the segmented query phrases as units of query representation. We investigate whether our method can improve three relevance models, namely BM25, key n-gram model, and dependency model. Our experimental results on three large scale web search datasets show that our method can indeed significantly improve relevance ranking in all the three cases.
\end{abstract}

\section{Introduction}\label{sec:intro}
Queries in web search are usually of three types: single phrases, combinations of phrases, and natural language questions, while natural language questions only consist of a small percentage. Traditionally, a query is viewed as a bag of words or a sequence of n-grams, and relevance models such as BM25~\citep{Robertson96}, dependency model~\citep{Metzler2005,Bendersky2011}, key-ngram model~\citep{Wang2012} utilize the words or n-grams as units of query representation.

A question naturally arises here. Is it possible to improve search relevance by conducting query segmentation first and then using the result as query representation in the relevance models? For example, if the query is ``my heart will go on mp3 download'', then one may want to segment the query into three segments: ``my heart will go on / mp3 / download". On the other hand, if the query is ``hot dog", then one may want to view it as a phrase rather than two separate words. The assumption is that the relevance can be improved by query segmentation in both cases.

Methods have been proposed for query segmentation. For example, \citet{Bergsma07} propose performing query segmentation by using a classifier. \citet{Hagen2011,Hagen2012} propose using unsupervised methods, more specifically, heuristic functions to conduct the task. Their methods outperform many existing methods and are viewed as state-of-the-art methods. For other methods see work of \citet{Jones2006,Tan2008,Zhang2009,Brenes2010,Huang2010,Risvik2003,Yu2009}.

Efforts have also been made for improving relevance ranking by using query segmentation. In most of the cases, the phrases in the segmented queries are directly used in the representations of queries. Most of the studies show that query segmentation is helpful, but either with quite small datasets~\citep{Bendersky2009,SahaRoy2012,Hagen2012} or using none standard relevance ranking models~\citep{Yanen2011}. There are also studies indicating that query segmentation does not help as expected~\citep{SahaRoy2012}.

In this paper, we study the problem of query segmentation for relevance ranking in web search. Our goal is to enhance the state-of-the-art methods for the task and to investigate the problem with large scale experiments.

We first propose a new method for query segmentation, on the basis of re-ranking, which is proved to be powerful for making structure prediction on sequence data in natural language processing, for example, part-of-speech tagging. The idea is to first pass the sequence and make prediction using a generative model to obtain the top $k$ candidates and then to rank the candidates to select the best result using a discriminative model. Although the approach is not new, it does not seem to have been applied to query segmentation. We consider a specific implementation of the approach, which uses ~\citet{Hagen2011}'s method (unsupervised learning) to find the top $k$ candidates and then uses SVM (supervised learning) to find the best segmentation among the candidates. We make comparison with the methods proposed by \citet{Hagen2011,Hagen2012} and \citet{Bergsma07}.

Next, we propose a new method for using query segmentation in search. In the method we take both the original query words and the query phrases obtained in segmentation as units and represent the query as bag of `units' or sequence of `units' (e.g., ``hot", ``dog", ``hot dog" are viewed as units) in the relevance models. We specifically construct BM25, key n-gram model \citep{Wang2012}, and dependency model \citep{Metzler2005,Bendersky2011} by using the query representation, respectively. We next take the scores of the models as features in the learning to rank model of LambdaMART~\citep{Burges2010} and employ the LambdaMART models in relevance ranking, respectively. We make comparison with the methods in which the same models are employed but no query segmentation is carried out, as well as the methods in which the same models are employed but only segmented query phrases are utilized (similar to some of previous work).

We make use of the benchmark datasets in previous work of query segmentation. The first dataset contains 500 queries and each query is segmented by three annotators~\citep{Bergsma07}. The second dataset contains 4850 queries and each query is segmented by ten annotators~\citep{Hagen2011}. We also use a large-scale dataset from a commercial web search engine for the experiments of relevance ranking. The dataset consists of 240,000 random queries, associated web pages (on average 43 pages for each query), and relevance judgments on the pages with respect to the queries.

Several conclusions have been made from the experimental results. (1) Our method of re-ranking in query segmentation works significantly better than the state-of-the-art methods on two public benchmark datasets. (2) Our method of using query segmentation in relevance ranking can help improve search relevance in all three cases in which the relevance schemes are BM25, key n-gram model, and dependency model. The improvements in terms nDCG are statistically significant. It is better to utilize both ngrams of original query words and ngrams of segmented query phrases in the relevance models.

The rest of the paper is organized as follows. Section~\ref{sec:related} introduces related work. Section~\ref{sec:segmentation} describes our method of query segmentation based on re-ranking. Section~\ref{sec:relevance} explains our method of using query segmentation in relevance ranking. Section~\ref{exp:segmentation},~\ref{exp:relevance} presents the experimental results on query segmentation as well as relevance ranking with query segmentation. And finally Section~\ref{sec:conclusions} concludes the paper.

\section{Related Work}\label{sec:related}

\subsection{Query Segmentation}
One of the earliest work for query segmentation is by \citet{Risvik2003}. They assume that a segment in a query should satisfy two conditions: it significantly frequently occurs in different sources and has a ``strong" mutual information. They calculate the likelihood of a segment by using the product of mutual information within the segment and the frequency of the segment in query log. \citet{Jones2006} also use mutual information to segment a query into phrases. They focus on two-word phrases and view any bigram whose mutual information is above a threshold as a phrase. \citet{Huang2010} make use of a web scale language model to address long query segmentation. They use mutual information scores obtained by the language model to create a segmentation tree and find the best segmentation by pruning the tree.  None of these methods is evaluated on a public dataset.

\citet{Bergsma07} have published the first benchmark dataset for research on query segmentation, referred to as ``Bergsma-Wang-Corpus" (BWC). The dataset contains 500 queries and each query is segmented by three annotators. They also add 500 labeled training queries and 500 labeled development queries for supervised learning approaches. The dataset is used in several previous work~\citep{Brenes2010,Hagen2011,Tan2008,Zhang2009,Hagen2012,Yanen2011,SahaRoy2012}. \citet{Hagen2011} have released a larger dataset ``Webis-QSec-10" (WQS) containing 4850 queries. The queries are sampled from the AOL query log, with segmentations annotated by ten annotators.

Both supervised and unsupervised methods have been developed for query segmentation. \citet{Bergsma07} propose a supervised method, which exploits a classifier using indicator features such as positions and POS tags, statistical features such as phrase frequencies on the web or in the query log, and dependency features such as frequencies of two adjacent words. \citet{Yu2009} employ a statistical method based on Conditional Random Fields (CRF) for which the parameters are learned from query logs. \citet{Bendersky2011b} propose jointly performing query segmentation, capitalization, and POS tagging on the query on the basis of CRF. \citet{Hagen2011,Hagen2012}, \citet{Tan2008} and \citet{Zhang2009} propose unsupervised methods, which employ heuristic functions and do not need to use data for training. Hagen et al. and Tan et al. make use of web n-grams from a large web corpus and titles of Wikipedia articles. Zhang et al. compute segment scores from the eigenvalues of the correlation matrix with regard to the given query.

\citet{Hagen2011,Hagen2012} report that their two methods outperform the existing unsupervised and supervised methods on the two public datasets Bergsma-Wang-Corpus and Webis-QSec-10-Corpus. We take their two methods as baselines in this paper.

\subsection{Relevance Ranking}
Relevance ranking is one of the key problems in web search. Given a query, documents containing the query words are retrieved, each document is assigned a score by the relevance model, and the documents are sorted on the basis of their scores. The relevance score of a document represents the relevance degree of the document with respect to the query. Traditionally, the relevance ranking model is manually created, with a few parameters to be tuned. For example, BM25~\citep{Robertson96}, Language Model for IR~\citep{Lafferty2001,Ponte98}, and dependency model~\citep{Metzler2005,Bendersky2010,Bendersky2011} are considered as the state-of-the-art schemes.

In the methods described above, n-gram (e.g., unigram) is usually used as a unit for calculating the relevance between query and document. In fact, the query and document can be represented as two n-gram vectors, and the relevance between them can be calculated as similarity between the two vectors (e.g., cosine similarity)~\citep{Xu10}. Intuitively, if the n-grams of the query occur more frequently in the document, then it is more likely that the document is relevant to the query. Methods have also been proposed to enhance relevance by conducting better n-gram based query document matching. For example, \citet{Wang2012} propose a method to extract key n-grams from the document (webpage) and then utilize the extracted key n-grams to augment the query document matching.

In web search, the title, anchor texts, URL, body, and associated queries of a web page can be used as multiple fields (pseudo texts) of the page in calculation of the relevance score~\citep{Wang2012}. Title, URL, and body are from the web page itself and reflect the author's view on the page. Anchor texts are from other pages and represent other authors' view on the page. Associated queries are from searchers and represent searchers' view on the page.

Recently, supervised learning techniques, called learning to rank, have also been proposed and have been proved to be useful in automatic combination of relevance models to create final ranking list of documents with respect to query, particularly in web search~\citep{Li2011b,Liu2009}. Among the learning to rank methods, the method of LambdaMART~\citep{Burges2010} is regarded as the state-of-the-art.

\subsection{Utilization of Query Segmentation Result}
Recently, the question of whether query segmentation can enhance relevance ranking has attracted researchers' interest. Several research groups have studied the problem~\citep{Bendersky2009,Hagen2012,SahaRoy2012,Yanen2011}. The conclusions from the investigations are not very consistent, however.

\citet{Bendersky2009} apply query segmentation into relevance ranking. Their goal is to investigate the effect of incorporating query segments into the dependency model based on Markov Random Fields (MRF)~\citep{Metzler2005}. They employ a linear model to combine the MRF scores of term matches, ordered segment matches, and unordered segment matches. Their experiments on different TREC collections indicate that query segmentation can indeed enhance the performance of relevance ranking. \citet{Yanen2011} have studied query segmentation in web search by exploiting click-through data. They incorporate query segments into language models using 1,000 queries. Experiments demonstrate the effectiveness of their method. There is difference between our method and the methods of \citet{Bendersky2009} and \citet{Yanen2011}. Both relevance ranking method and query segmentation method are supervised in our case, while their methods are all unsupervised.

\citet{Hagen2012} and \citet{SahaRoy2012} have studied whether modifying queries by adding quotations of phrases into them can improve relevance ranking at a web search engine. Note that using quotations is nearly equivalent to utilizing segmented queries\footnote{In practice, less than 1.12\% queries include quotations~\citep{White07}.}. They submit the results of query segmentations by different methods to the search engine. Their results show that in most cases query segmentation can help generate better relevance ranking. However, sometimes it is better not to conduct query segmentation. They only treat the search engine as a black box and do not make use of the results of query segmentation inside the search engine.

\section{Our Method of Query Segmentation}\label{sec:segmentation}
\subsection{Problem}
Query segmentation is to separate the query into disjoint segments so that each segment roughly corresponds to a phrase (note that it is not necessarily to be a phrase in natural language.). Given a query $Q=w_1,w_2,\cdots,w_n$ of length $n$ where $w_i, i=1,\cdots,n$ denotes a word. A segmentation of $Q$ is represented as $S=s_1s_2\cdots s_m$ of length $m$ where $s_i, i=1,\cdots,m$ denotes a segment. There are $2^{n-1}$ possible segmentations and $(n^2+n)/2$ possible segments for $Q$. Therefore, query segmentation is equivalent to selecting the best segmentation from among the possible ones given the query.

For convenience, we sometimes use breaks (boundaries between a pair of adjacent words) to represent a segmentation. A segmentation can also be represented as $B=b_1b_2\cdots b_{(n-1)}$ of length $n-1$ where $b_i \in \{1,0\}, i=1,\cdots,n-1$ denotes a break, 1 stands for making the break, and 0 stands for not making the break. There are $n-1$ breaks for query $Q$ of length $n$.

\subsection{Method}\label{subsec:rerank}

We take a re-ranking approach to query segmentation. First, we employ \citet{Hagen2011}'s unsupervised method to find the top $k$ candidate segmentations, and then employ the supervised learning method of SVM to re-rank the candidates and find the best segmentation as output. Dynamic program is applied to generate the top $k$ segmentations with the highest scores, which reduces the time complexity from $O(2^n)$ to $O(n^2)$ where $n$ denotes the query length. The two-stage approach is widely used in other tasks in natural language processing. To the best of our knowledge, this is the first time it is used for query segmentation.

\citet{Hagen2011} propose a method called Wikipedia-Based Normalization(WBN) which is the state-of-the-art on the most widely used benchmark dataset~\citep{Bergsma07}. Given a segmentation, WBN assigns a weight to each segment and sum up all the weights as the score of the entire segmentation. We choose the segmentations with the highest $k$ scores. The score of segmentation $S$ is defined as below:

\begin{equation}
    \nonumber
    \begin{aligned}
        score(S)=
        \left\{
            \begin{array}{cl}
                {\sum\limits_{s\in S, |s|\geq 2}weight(s)}    &\begin{split}&\text{if } weight(s) > 0 \text{ for } \\& \text{all } s\in S \text{ and }|s|\geq 2
                 \end{split} \\\\
                -1 & \text{else.}
            \end{array}
        \right.
    \end{aligned}
\end{equation}
        where $s$ is a segment and segments with length one is ignored. The weight of segment $s$ is defined as below,
\begin{equation}
        \nonumber
        weight(s)=
        \left\{
            \begin{array}{ll}
                |s|^2+|s|\cdot{\max\limits_{t\in s, |t|=2}}~freq(t)    & \begin{split}&\text{if } s \text{ is Wikipedia} \\&\text{title}
                                                                        \end{split} \\\\
                |s|\cdot freq(s) & \text{else.}
            \end{array}
        \right.
\end{equation}
where $t$ denotes a substring of $s$ and $freq$ denotes the frequency of string in the corpus.

There are also other approaches that can achieve good results in query segmentation. For example, \citet{Bergsma07}'s learning-to-rank method which generates short segments by focusing on noun phrases and leaving the rest as one-word segments. \citet{Hagen2012}'s another heuristic method based on Wikipedia titles makes a slight modification from WBN, which also generates short segments by highlighting Wikipedia title segments and leaving the rest as one-word segments. We choose \citet{Hagen2011}'s WBN method in this paper, since it performs the best on the most widely used benchmark dataset published in \citep{Bergsma07}.

We investigate the top ranked segmentations by the WBN method, and find that for a given query the probability of finding the ground truth appearing in the top six ranked segmentations is 94\%. Therefore, we only select the top six segmentations in our experiments. (See details in Section~\ref{subsec:para})

We adopt SVM~\citep{joachims2002} as the method to train the re-ranking model. We take the segmentations in the ground truth as positive examples, and the other segmentations among the top six segmentations as negative examples. Training data is then constructed, and Table~\ref{tab:example_rerank} gives an example. The re-ranking model is trained with the training data.

\begin{table}[t]
\caption{\label{tab:example_rerank} Examples of positive instance and negative instances. The segmentation consistent with human label is regarded as positive, and others are regarded as negative.}
\centering
\linespread{1.1}\selectfont
\begin{center}
\centering
\(
\begin{tabular}{clcc}\thickhline
\multirowcell{2}{\textbf{No.}} &\multirowcell{2}{\textbf{Segmentation}}           &\multirowcell{2}{\textbf{Human}\\ \textbf{Label}}      &\multirowcell{2}{\textbf{Label}}  \\
&&&\\\hline
1    &beijing / seven eleven stores     &$\times$   &-1     \\
2    &beijing / seven eleven / stores   &$\bigcirc$ &+1     \\
3    &beijing seven eleven / stores   &$\times$   &-1     \\
4    &beijing seven / eleven stores   &$\times$   &-1     \\
5    &beijing / seven / eleven stores   &$\times$   &-1     \\
6    &beijing seven / eleven / stores &$\times$   &-1     \\\thickhline
\end{tabular}
\)
\end{center}
\end{table}

\subsection{ Features of Re-ranking Model}

Table~\ref{tab:features} show the features used in the re-ranking method.

The first group of features are those obtained from the ranking function of Hagen et al.'s  Wikipedia-based normalization method described in Section~\ref{subsec:rerank}. See features $F(1,\cdot)$ in Table~\ref{tab:features} .

The second group of features utilize mutual information, which has been proved to be useful for query segmentation~\citep{Risvik2003,Jones2006,Huang2010}. The assumption is that a good segmentation should have low MI values between segments and high MI values within a segment. See features $F(2,\cdot)$ in Table~\ref{tab:features}.

The third group of features represent the characteristics of segmentations of query. All the features are of Boolean type. See features $F(3,\cdot)$ in Table~\ref{tab:features}.

We observe that the top candidate segmentations with the highest scores tend to be close to the ground truth. The four group of features consider the similarities between the current segmentation and the top segmentation.

Suppose that the current segmentation is $S$ ($B=b_1b_2\cdots b_{n-1}$) and the top segmentation is $S^h$ ($B^h={b_1^hb_2^h\cdots b_{n-1}^h}$). We measure the similarities between the two segmentations in several ways. See features $F(4,\cdot)$ in Table~\ref{tab:features}.

\begin{table}[th]
\caption{\label{tab:features} Features of re-ranking model for query segmentation}
\footnotesize
\centering
\linespread{1.1}\selectfont
\begin{center}
\centering
\(
\begin{tabular}{l|b{32em}}
	\thickhline
 \textbf{Feature}                          & \textbf{Description}           \\\hline
$F(1,1)$& the rank of the segmentation.	\\\hline
$F(1,2)$& the score of the segmentation. \\\hline
$F(1,3)$&the sums of weights of segments in different lengths. There are six such features, corresponding to segment lengths of 1, 2, 3, 4, 5 and beyond.\\\hline
$F(1,4)$& the weight of the first segment.\\\hline
$F(1,5)$& the average weight of the segments.\\\hline
$F(1,6)$& the number of the segments.\\\hline
$F(1,7)$& the average length of the segments.\\\hline
$F(1,8)$& the maximum length of the segments which are Wikipedia titles.\\\hline\hline
$F(2,1)$& the maximum value of mutual information of all the adjacent segments. If the segmentation has only one segment, this feature equals zero.\\\hline
$F(2,2)$& the maximum value of mutual information of the adjacent words. If the segmentation has only one segment, this feature equals zero.\\\hline
$F(2,3)$& the minimum value of mutual information of the adjacent words in a segment. The segment with only one word is not considered here.\\\hline\hline
$F(3,1)$& There are words which tend to form a single word segment, such as ``and", ``vs". There are eighteen such words. If one of them occurs, then the value of the corresponding feature becomes one.\\\hline
$F(3,2)$& If the first (or last) segment has two words, then the feature value becomes one.\\\hline
$F(3,3)$& If there is a subsequence of words with their first letters capitalized, then the feature value becomes one.\\\hline
$F(3,4)$& If one segment is multi-word segment and the other segments consist of only one word in the segmetation, then the feature  value becomes one.\\\hline
$F(3,5)$& the number of one-word segments in the segmentation.\\\hline\hline
$F(4,1)$& If splitting one segment in $S^h$ will make it equivalent to $S$, then the feature value is one. For $i$, $b_i=1$ and $b_i^h=0$ hold and for $j (j\neq i)$, $b_j=b_j^h$ holds.\\\hline
$F(4,2)$& If merging two segments into one in $S^h$ will make it equivalent to $S$, then the feature value is one. For $i$, $b_i=0$ and $b_i^h=1$ hold and for $j (j\neq i)$, $b_j=b_j^h$ holds.\\\hline
$F(4,3)$& If moving a break from one place to the other in $S^h$ will make it equivalent to $S$, then the feature value becomes one. For $i$, $b_ib_{i+1}=01 \text{ }(\text{or }10)$ and $b_i^hb_{i+1}^h=10 \text{ }(\text{or }01)$ hold and for $j (j\neq i, i+1)$, $b_j=b_j^h $ holds.\\\hline
$F(4,4)$& the number of identical breaks with the top segmentation.\\\hline
$F(4,5)$& the number of identical segments with the top segmentation.\\
	\thickhline
\end{tabular}
\)
\end{center}
\end{table}

\section{Our Method of Relevance Ranking}\label{sec:relevance}

We propose a new method for employing query segmentation in relevance ranking. It seems to be new, as far as we know, although the idea is quite simple.

\subsection{General Principle}\label{subsec:rankingfeatures}
In principle, using segmented queries is useful for finding relevant information. For example, when the query is ``china kong movies description"\footnote{China Kong is an American actor and a producer in 1980s.}, it is better to segment the query to ``china kong / movies / description" and take china kong as a unit in the query representation. On the other hand, there is no guarantee that query segmentation can be performed perfectly, no matter which method is employed. It would also be necessary to retain the original query words in the query representation.

Our approach makes use of both query words and segmented query phrases as units of query representation and employs the query representation in the relevance ranking model. More specifically, given a query, our approach conducts segmentation on the query by a query segmentation method, for example, our method described in Section~\ref{sec:segmentation}. It then builds two feature vectors to represent the query. The first vector is comprised of the query words and the second vector is comprised of the segmented query phrases.  The two feature vectors are then utilized in the relevance ranking model. If we do not use the second vector, then our approach degenerates to the approach of not using query segmentation. If we do not use the first vector, then our approach becomes equivalent to the existing methods of only using segmented query phrases.

For example suppose that the query is ``beijing seven eleven stores" and it is segmented into ``beijing", ``seven eleven", ``stores" by a segmentation method. Our approach uses ``beijing", ``seven", ``eleven", ``stores" to represent the first feature vector in which the corresponding elements are one and the other elements are zero.  It further uses ``beijing", ``seven eleven", ``stores"  to represent the second feature vector in which the corresponding elements are one and the other elements are zero. This is for unigrams, and we can easily extend to bigrams and trigrams, as shown in Table~\ref{tab:example_seg_ngram}.

\begin{table*}[t]
\caption{\label{tab:example_seg_ngram} Example of query representation.}
\centering
\footnotesize
\linespread{1.1}\selectfont
\begin{center}
\centering
\(
\begin{tabular}{l|p{15em}|p{15em}}\thickhline
            &\textbf{Query Word Based}                                          &\textbf{Query Phrase Based}                                   \\
            &beijing seven eleven stores                    &bejing / seven eleven / stores               \\\hline
Unigram     &beijing, seven, eleven, stores                 &beijing, seven eleven, stores                \\\hline
Bigram      &beijing seven, seven eleven, eleven stores     &beijing seven eleven, seven eleven stores   \\\hline
Trigram     &beijing seven eleven, seven eleven stores      &beijing seven eleven stores                 \\\thickhline
\end{tabular}
\)
\end{center}
\end{table*}

\subsection{Method}

We describe how to employ our approach when the relevance ranking scheme is BM25~\citep{Robertson96}, key n-gram model~\citep{Wang2012}, and dependency model~\citep{Bendersky2011}, three state-of-the-art models, respectively.

In web search, web pages (documents) are represented in several fields. We consider the use of the following fields: URL, title, body, meta-keywords, meta-description, anchor texts and associated queries in search log data. Each document is represented by its fields and is indexed in the search system. In search, BM25 model, key n-gram model or dependency model is created for each field of document.

We consider n-gram BM25 (simply referred to as BM25), which is a natural extension of the traditional BM25 based on unigrams. Given the query representation described above as well as the document representation in the index, we calculate the n-gram BM25 score for each field of the document with respect to each feature vector of the query. Therefore, each field has six BM25 scores  calculated based on word based unigrams, word based bigrams, word based trigrams, phrase based unigrams, phase based bigrams, and phrase based trigrams, respectively.  To calculate a BM25 score we need to use the term frequencies of n-grams, document frequencies of n-grams, number of documents, and document length. The first three numbers can be easily obtained, but the last one can only be estimated since the document is not segmented. We use the traditional n-gram document length~\citep{Wang2012} to approximate the document length. Finally, we employ LambdaMART~\citep{Burges2010} to automatically construct the ranking model with all the n-gram BM25 scores of all the fields as features. Since there are seven fields and each field has six BM25 features, there are in total forty-two features in the ranking model.

When exploiting the key n-gram  scheme, we extract key unigrams, bigrams, and trigrams from the body of the web page and create an additional stream with all the extracted key n-grams combined together, in the same way as \citet{Wang2012}. We then calculate the n-gram BM25 scores for all the fields including the key n-gram field, similarly to the BM25 model above. We employ LambdaMART~\citep{Burges2010} to automatically build the final ranking model with all the BM25 scores of all the fields as features, as proposed by \citet{Wang2012}. There are in total forty-eight features.

When exploiting the dependency scheme, we only make use of unigrams and bigrams in query representation, following the practice as \citet{Bendersky2011}. Each unigram or bigram has seven weights calculated by using other data sources such as web n-gram, query log, and Wikipedia. Each unigram is assigned with the normalized frequency of it in a field of the document, and each bigram is assigned with the normalized frequency of its consecutive occurrences in a field of the document and the normalized frequency of its inconsecutive occurrences within a window of size eight in a field of the document. The product of weight and normalized frequency of a unigram or bigram is calculated. The sums of weighed normalized frequencies are calculated over the unigrams and bigrams and they are taken as features of unigrams and bigrams, respectively. Since there are seven weights and three normalized frequencies, there are twenty-one features for each field (URL, title, body, meta-keywords, meta-description, anchor texts and associated queries) and each query vector (query word based and query phase based). We again employ LambdaMART~\citep{Burges2010} to automatically construct the final ranking model, which is similar to the coordinate descent method utilized by \citet{Bendersky2011}. In total, there are two hundred and ninety-four features in the model.

\section{Experiments on Query Segmentation}\label{exp:segmentation}
In this section, we report the experimental results of query segmentation.
\subsection{Experiment Setup}
\subsubsection{Datasets}
We use two public datasets in our experiments: \textbf{Bergsma-Wang-Corpus} (BWC)~\citep{Bergsma07} and \textbf{Webis-QSec-10 Corpus} (WQS)~\citep{Hagen2011}. BWC consists of 500 queries sampled from the AOL query log dataset~\citep{Pass06}. The queries only contain determiners, adjectives, and nouns, and have a length of four or greater. Each query has three segmentations labeled by three annotators. WQS consists of 4850 queries randomly sampled from the AOL the query log dataset, and each query is labeled by ten annotators.

In BWC (500 queries), there are 220 queries (44\%) for which the three annotators have an agreement, and there are 454 queries (91\%) for which at least two of the three annotators have an agreement. In WQS (4850 queries), there are only 167 queries (3.4\%) for which all the ten annotators have an agreement, and there are 3769 queries (78\%) for which half of the annotators have an agreement.

\citet{Hagen2012} propose a break fusion method for determining the gold standard of a dataset labeled by multi labelers. We adopt the method, since it is reasonable and easy to implement. For each position between a pair of adjacent words, if at least half of the labelers insert a break, then the method also inserts a break.

Table~\ref{tab:segdist} shows the distributions of segments in different lengths of the two datasets, as well as the average segment lengths. Notice that BWC favors longer segments, while WQS favors shorter segments.

\begin{table}[h]
\caption{\label{tab:segdist} Distributions of segments in different lengths in two datasets. }
\centering
\small
\linespread{1.1}\selectfont
\begin{center}
\centering
\(
\begin{tabular}{lccccccc}
	\thickhline
\multirow{2}{*}{\textbf{Dataset}} & \multirow{2}{*}{\textbf{Query \#}} & \multirowcell{2}{\textbf{Word \#}\\\textbf{per query}} &\multicolumn{4}{c}{\textbf{Segment length ratio}} & \multirowcell{2}{\textbf{Word \#}\\\textbf{per segment}}\\
	                                              &       &                                &  1   &  2   &               3                & 4$+$   \\ \hline
	BWC                                           &  500  &              4.3               & 32\% & 55\% &              9\%               & 4\%  & 1.9   \\
	WQS                                           & 4,850 &              4.1               & 67\% & 26\% &              6\%               & 1\%  & 1.4   \\
	\thickhline
\end{tabular}
\)
\end{center}
\end{table}

\subsubsection{Evaluation Measures}
There are five widely used measures for evaluation of the performance of a query segmentation method~\citep{Hagen2011}: {\em segmentation accuracy} stands for the ratio of segmented queries exactly matching with the ground truth, {\em segment precision} stands for the ratio of correct segments among all generated segments, {\em segment recall} stands for the ratio of correct segments among all segments in the ground truth, {\em segment F-Measure} stands for the harmonic mean of the former two measures, and {\em break accuracy} stands for the ratio of correct breaks between two adjacent words.

\subsubsection{Baselines}
As baselines, we choose two state-of-art methods by Hagen et al. and one of the earliest methods by Bergsma et al.: the Wikipedia-Based Normalization method~\citep{Hagen2011} (denoted as WBN), the Wikipedia-Title method~\citep{Hagen2012} (denoted as WT) and the Noun-Phrase method~\citep{Bergsma07} (denoted as NP). WBN and WT had the best performances on the BWC and WQS datasets respectively. In our implementation of the methods, we use the Microsoft Web $N$-Gram Servcie\footnote{\url{http://research.microsoft.com/en-us/collaboration/focus/cs/web-ngram.aspx}} to calculate the web $n$-gram frequencies and query $n$-gram frequencies, use the Wikipedia database\footnote{\url{http://en.wikipedia.org/wiki/Wikipedia:Database\_download}} to decide whether an n-gram matches with a Wikipedia title, and use Stanford Parser\footnote{\url{http://nlp.stanford.edu/software/tagger.shtml}} to collect POS information.

\subsubsection{Parameter Tuning}\label{subsec:para}
No parameter needs to be tuned in the unsupervised methods of WBN and WT. There are three parameters to be tuned in NP and four parameters to be tuned in our method.

Both the NP method and our method use SVM to train the model for query segmentation, and the tool we use is SVMLight~\citep{joachims2002}. There are three parameters $\{c,j,b\}$\footnote{See \url{http://svmlight.joachims.org/} for the formal definitions of the parameters}. We set the range of $c$ as $\{0.01, 0.02, 0.05, 0.1,\cdots, 20, 50\}$, the range of $j$ as $\{1, 1.5, 2, 2.5,\ldots, 8\}$, and the range of $b$ as $\{1, 0\}$. We conduct four-fold cross validation to choose the best parameter settings for our method and NP with respect to the two datasets, while taking segmentation accuracy as evaluation measure. We find that the best settings for NP are $\{0.1,1,0\}$ for BWC and $\{0.05,1,1\}$ for WQS, and the best settings for our method are $\{2,1,1\}$ for BWC and $\{20,2,0\}$ for WQS.

There is one more parameter $k$ for our method for selecting the top $k$ segmentations. Table~\ref{tab:top_seg} shows the probability of the correct segmentations appearing among the top $k$ candidate segmentations by WBN. We can see that the probability reaches 0.94, when $k$ is 6. Thus, we choose $k=6$ to make a good trade-off between accuracy and efficiency in our experiments.

\begin{table}[t]
\caption{\label{tab:top_seg} The probability of correct segmentation appearing in the top $k$ candidate segmentations by the method of WBN for the BWC dataset. }
\centering
\linespread{1.1}\selectfont
\begin{center}
\centering
\(
\begin{tabular}{lccccccc}
\thickhline
\textbf{Top $N$} & \textbf{1}     & \textbf{2}     & \textbf{3}     & \textbf{4}     & \textbf{6}     & \textbf{8}     & \textbf{10}    \\\hline
Prob.   & 0.50  & 0.73  & 0.81  & 0.87  & 0.94  & 0.98  & 0.98  \\\thickhline
\end{tabular}
\)
\end{center}
\end{table}

\subsection{Results and Discussions}\label{subsec:qs_results}
We compare the effectiveness of our method and the three baselines on query segmentation using the two datasets.

Table~\ref{tab:qs_results} shows the results on the two datasets in terms of the five measures. The results of three baselines are comparable with those reported by \citet{Hagen2012}. It is evident that our method of re-ranking outperforms the baselines of WBN and WT in terms of all measures except segment recall\footnote{This is because the NP method tends to generate shorter segments~\citep{SahaRoy2012}, while most human labeled segments are shorter than 3 words (See Table~\ref{tab:segdist})}. Especially on the WQS dataset, all the improvements are statistically significant on sign-test ($p<0.01$). The result demonstrates that our method is indeed effective and can enhance the accuracy of query segmentation.

\begin{table}[t]
    \caption{\label{tab:qs_results} Our method of re-ranking consistently makes improvements upon the baselines on the BWC and WQS datasets in terms of all measures. }
    \centering
    \linespread{1.1}\selectfont
    \begin{center}
    \centering
    \(
        \begin{tabular}{llllll}
        	\thickhline
        \multirow{2}{*}{\textbf{Corpus}} & \multirowcell{2}{\textbf{Performance}   \\
        	\textbf{Measure}}                                    &                                  \multicolumn{4}{c}{\textbf{Algorithm}}                                    \\
        	                                                     &                                       & NP                  & WT                  & WBN                  & Our             \\ \hline
        	\multirowcell{5}{BWC}                                & query acc                             & 0.548               & 0.414               & 0.572                & \textbf{0.602}*  \\
        	                                                     & seg prec                              & 0.651               & 0.538               & 0.692                & \textbf{0.715}*  \\
        	                                                     & seg rec                               & \textbf{0.742}      & 0.658               & 0.664                & 0.700           \\
        	                                                     & seg F                                 & 0.694               & 0.592               & 0.677                & \textbf{0.707}*  \\
        	                                                     & break acc                             & 0.834               & 0.762               & 0.830                & \textbf{0.848}*  \\ \hline
        	\multirowcell{5}{WQS}                                & query acc                             & 0.512               & 0.508               & 0.362                & \textbf{0.560}* \\
        	                                                     & seg prec                              & 0.666               & 0.680               & 0.561                & \textbf{0.710}* \\
        	                                                     & seg rec                               & \textbf{0.796}      & 0.728               & 0.456                & 0.749           \\
        	                                                     & seg F                                 & 0.726               & 0.703               & 0.503                & \textbf{0.729}  \\
        	                                                     & break acc                             & 0.783               & 0.784               & 0.680                & \textbf{0.800}* \\
        	\thickhline
        \end{tabular}
    \)\\
    \textbf{Bold}: the maximum value of the performance measure.\\
    *: statistically significant improvement from all baselines (sign-test, $p<0.01$).
    \end{center}

\end{table}

We examine the weights of the linear SVM model in our method, in which a higher weight indicates a more important contribution. First, the ``rank" and ``score" features have the highest weights, which can ensure that the re-ranking method has similar performances as WBN and WT. Besides, the features of ``weight on segment length" also have high weights. The features can capture the tendencies of segment lengths in different datasets and thus can help improve the performances in different datasets. (Recall that in BWC and WQS segments with different lengths are preferred.)

In addition, our method of re-ranking can leverage information which WBN and WT cannot, such as mutual information. For example, for query ``play disney channel games", both WBN and WT treat ``disney channel games" as a segment, since it is also a Wikipedia title. However, it seems that the user is searching for games on the Disney Channel instead of searching for the Wikipedia page. (In fact, there is a webpage entitled ``Games | Disney Channel" which can perfectly meet the need.) Therefore, the annotators label the query as ``play / disney channel / games". The feature of ``min MI of words in segment" can help our method to effectively deal with the problem. The adjacent words ``channel games" has a small mutual information value, indicating that they should be separated. This is the main reason that our method can work better than the baselines.

\section{Experiments on Relevance Ranking}\label{exp:relevance}
In this section, we report the experimental results of relevance ranking using query segmentation.

\subsection{Experiment Setup}
\subsubsection{Dataset}
We conduct experiments on relevance ranking using a large data set collected from a commercial search engine. The data set contains queries, documents, and relevance judgments. The relevance judgments are represented at five levels including ``Perfect(4)", ``Excellent(3)", ``Good(2)", ``Fair(1)", and ``Bad(0)". The relevance judgment of each query-document pair is the average relevance score by three labelers. The whole data set is split into five subsets: Training data set for learning, Validation data set for parameter tuning, and Test1, Test2, Test3 data sets for evaluation. Training, Validation and Test3 are comprised of general queries (randomly sampled from search log), associated documents and their relevance judgments. Test1 consists of head queries (with high frequencies and randomly sampled from the search log), associated documents, and their relevance judgments. Test2 consists of tail queries (with low frequencies and randomly sampled from the search log), associated documents, and their relevance judgments. The subsets do not have any overlap with each other. Statistics on the dataset is given in Table~\ref{tab:rankds}.

\subsubsection{Evaluation Measure}
To evaluate the relevance performance of different ranking models, we calculate Normalized Discounted Cumulative Gain (NDCG)~\citep{Jarvelin2000} at positions 1, 5 and 10.

\begin{table}[t]
\caption{\label{tab:rankds} Dataset in relevance ranking.}
\centering
\linespread{1.1}\selectfont
\begin{center}
\centering
\(
\begin{tabular}{l|rrrrr}
	\thickhline
\multirow{2}{*}{\textbf{Data}} & \multirowcell{2}{\textbf{Training}\\\textbf{(random)}} & \multirowcell{2}{\textbf{Validation}\\\textbf{(random)}} & \multirowcell{2}{\textbf{Test1}\\\textbf{(head)}} & \multirowcell{2}{\textbf{Test2}\\\textbf{(tail)}} & \multirowcell{2}{\textbf{Test3}\\\textbf{(random)}} \\\\ \hline
	Queries \#                & 201,585        & 3,953               & 12,089         & 10,490         & 10,959         \\
	HTML Pages \#             & 8,761,343      & 158,837             & 664,362        & 283,956        & 453,155        \\
	Pages \#/query            & 43.46          & 40.18               & 54.96          & 27.07          & 41.35          \\
	Perfect \#/query          & 0.24           & 0.26                & 0.33           & 0.12           & 0.25           \\
	Excellent \#/query        & 0.72           & 0.72                & 0.99           & 0.54           & 0.70           \\
	Good \#/query             & 5.61           & 5.15                & 9.93           & 4.14           & 5.32           \\
	Fair \#/query             & 12.71          & 12.59               & 20.89          & 9.05           & 12.65          \\
	Bad \#/query              & 23.16          & 21.46               & 22.82          & 13.22          & 22.43          \\
	Words \#/query            & 3.70           & 3.76                & 3.05           & 4.49           & 3.70           \\
	\thickhline
\end{tabular}
\)
\end{center}
\end{table}

\subsubsection{Our Approach and Baselines}

We use seven fields of a web page (document): url, title, body, meta-keywords, meta-description, anchor texts, and associated queries in search log. In the key n-gram model (KN), there is an additional key n-gram field.

We test the effectiveness of our approach of using query segmentation in three relevance ranking schemes: BM25, key n-gram model, and dependency model. (1) BM25: n-gram BM25 is utilized, where n-gram includes unigram, bigram, and trigram, denoted as BM25. (2) Key n-gram model: the key n-gram model of \citet{Wang2012} is employed, where n-gram includes unigram, bigram, and trigram, denoted as KN. (3) Dependency model: the dependency model is employed which is similar to the method of \citet{Bendersky2010}, denoted as DM.

Six query segmentations methods are considered in our approach. (1) The NP method~\citep{Bergsma07} trained with the BWC dataset, (2) The NP method trained with the WQA dataset. (3) The WBN method (no training is needed), (4) the WT method (no training is needed), (5) our re-ranking method trained with the BWC dataset, and (6) our re-ranking method trained with the WQS dataset. They are denoted as NP@BWC, NP@WQS, WBN, WT, RR@BWC, and RR@WQS.

We consider three baselines: BM25, KN, and DM, in which no query segmentation is employed. In other words, only query words are used in the relevance ranking schemes of BM2, key n-gram model, and dependency model.

\subsubsection{Parameter Tuning}
We use LambdaMART to train different gradient boosted trees as relevance ranking models. There are four parameters in LambdaMART: $\{nt, nl, lr, mil\}$, which stands for number of trees, number of leaves, learning rate, and minimum instances per leaf, respectively. We choose $nt$ from $\{10,50,100\}$, $nl$ from$\{2,4,16\}$, $lr$ from $\{0.1,0.2,0.3,0.4\}$, and $mil$ from $\{10,50,100\}$ for each ranking model using the Validation data.

\subsection{Main Results}

Table~\ref{tab:rr_seg} shows the results in terms of NDCG on Test1 (head queries), Test2 (tail queries) and Test3 (random queries). We use the following notations. For example, BM25-RR@BWC-WP means that the relevance scheme is BM25, segmentation method is RR@BWC, and both query words and query phrases are utilized. BM25-RR@WQS-WP means that the relevance scheme is BM25, segmentation method is RR@WQS, and both query words and query phrases are utilized.

\begin{table}[t]
\centering
\caption{\label{tab:rr_seg}The results on relevance ranking in three ranking schemes with six segmentation methods.}
\linespread{1.1}\selectfont
\begin{center}
\centering
\(
\scriptsize
    \begin{tabular}{p{9.7em}|p{3.2em}<{\centering}p{4.2em}<{\centering}|b{3.2em}<{\centering}p{4.2em}<{\centering}|b{3.2em}<{\centering}p{4.2em}<{\centering}}
    	\thickhline    & \multicolumn{2}{c|}{\textbf{Test1(head)}} & \multicolumn{2}{c|}{\textbf{Test2(tail)}} & \multicolumn{2}{c}{\textbf{Test3(random)}} \\
    	               & NDCG@5          & NDCG@10                 & NDCG@5          & NDCG@10                 & NDCG@5          & NDCG@10                  \\ \hline
    	BM25           & 54.27           & 56.62                   & 38.42           & 41.82                   & 44.66           & 47.07                    \\
    	BM25-NP@WQS-WP & 54.47           & 56.70                   & 38.42           & 41.83                   & 44.76           & 47.12                    \\
    	BM25-NP@BWC-WP & 54.60*          & 57.00*                  & 38.62           & 41.84                   & 44.86           & 47.20                    \\
    	BM25-WT-WP     & 54.64*          & 57.13*                  & 38.54           & 42.03                   & 44.88           & 47.21                    \\
    	BM25-WBN-WP    & 54.93*          & 57.62*                  & 38.73           & 42.13*                  & 45.06*          & 47.48*                   \\
    	BM25-RR@WQS-WP & 55.24*          & 57.75*                  & 39.16*          & 42.39*                  & 45.15*          & 47.48*                   \\
    	BM25-RR@BWC-WP & \textbf{55.82}* & \textbf{58.20}*         & \textbf{39.21}* & \textbf{42.45}*         & \textbf{45.22}* & \textbf{47.50}*          \\ \hline
    	KN             & 55.78           & 58.38                   & 40.23           & 43.90                   & 46.87           & 49.20                    \\
    	KN-NP@WQS-WP   & 55.82           & 58.44                   & 40.25           & 43.87                   & 47.07           & 49.50*                   \\
    	KN-NP@BWC-WP   & 56.00*          & 58.68*                  & 40.30           & 43.85                   & 47.12           & 49.56*                   \\
    	KN-WT-WP       & 57.09*          & 59.74*                  & \textbf{40.67}* & \textbf{44.07}          & 48.04*          & 50.29*                   \\
    	KN-WBN-WP      & 56.15*          & 56.95*                  & 40.33           & 43.99                   & 47.74*          & 49.92*                   \\
    	KN-RR@WQS-WP   & 56.33*          & 59.06*                  & 40.55*          & 43.99                   & 47.86*          & 50.16*                   \\
    	KN-RR@BWC-WP   & \textbf{57.71}* & \textbf{60.39}*         & 40.55*          & \textbf{44.07}          & \textbf{48.09}* & \textbf{50.37}*          \\ \hline
    	DM             & 57.71           & 60.16                   & 37.82           & 41.25                   & 47.80           & 50.07                    \\
    	DM-NP@WQS-WP   & 58.01*          & 60.56*                  & 37.84           & 41.30                   & 47.91           & 50.17                    \\
    	DM-NP@BWC-WP   & 58.83*          & 61.15*                  & 37.95           & 41.45                   & 47.98           & 50.31*                   \\
    	DM-WT-WP       & 59.16*          & 61.62*                  & 38.00           & 41.39                   & 48.62*          & 50.69*                   \\
    	DM-WBN-WP      & 59.13*          & 61.74*                  & 38.04           & 41.35                   & 48.76*          & 50.74*                   \\
    	DM-RR@WQS-WP   & 59.69*          & 62.01*                  & 38.16*          & 41.41                   & 48.89*          & 50.90*                   \\
    	DM-RR@BWC-WP   & \textbf{60.14}* & \textbf{62.54}*         & \textbf{38.28}* & \textbf{41.73}*         & \textbf{48.98}* & \textbf{51.04}*          \\ \hline
    	\thickhline
    \end{tabular}
\)\\
\textbf{Bold}: the highest performance for the scheme with respect to the dataset. \\
*: statistically significant improvement on baseline (t-test, $p<0.01$).
\end{center}
\end{table}

The experimental results show that the performances of all three schemes are improved in terms of all measures when our approach is employed. And most of the improvements are statistically significant by t-test ($p<0.01$), especially on the Test1 and Test3 Data. The results indicate that our approach of employing query segmentation is effective for relevance ranking.

We investigate the main reasons of the performance enhancement by our approach.

(1) The longest n-gram in BM25 and KN is trigram and that in DM is bigram. Thus, the models do not directly handle phrases longer than three words. For example, in query ``my heart will go on mp3 download", ``my heart will go on" is a phrase, and is not taken as a whole by the three baseline models. In contrast, there is no length constraint in query segmentation, and the query can be segmented into ``my heart will go on / mp3 / download". Our approach based on query segmentation can properly use the segmentation in relevance ranking.  This seems to be the main reason of the performance improvements by our approach.

(2) The baseline models put equal weights on the n-grams of the same lengths. In fact, some n-grams should have higher weights because they are of more importance in queries. Query segmentation can help to reduce the impact of meaningless n-grams and enhance the impact of meaningful n-grams. For example, for query ``beijing / seven eleven / store", segmentation can filter out meaningless bigrams ``beijing seven" and ``seven stores", which may have negative influence on relevance, and can retain meaningful bigrams such as ``seven eleven", which may have positive influence on relevance.

(3) In DM, only the dependencies between word bigrams are considered. Thus, it is not possible for DM to handle dependencies between phrases. In contrast, our approach based on query segmentation can cope with the problem, when there exist dependencies between phrases. For example, the dependency between phrases ``north korea" and ``nuclear weapon" in the query of ``North Korea nuclear weapon in 2009" is useful for relevance ranking, and can be leveraged by our approach.

\subsection{Comparison with Using Only Segmented Query Phrases}

The key idea of our approach is to make use of both query words and query phrases (i.e., create two query vectors). This is also one of the major differences between our approach and existing methods.

In this section, we make comparison between our approach and the alternative approaches which only use segmented query phrases, again in the three schemes.

Table~\ref{tab:rr_onlySeg} shows the results in terms of NDCG. We use the following notations. For example, BM25-RR@BWC-P means that the relevance scheme is BM25, segmentation method is RR@BWC, and only query phrases are utilized. BM25-RR@WQS-P means that the relevance scheme is BM25, segmentation method is RR@WQS, and only query phrases are utilized.

\begin{table}[t]
\centering
\caption{\label{tab:rr_onlySeg}The results on relevance ranking when only segmented query phrases are used as representation.}
\linespread{1.1}\selectfont
\begin{center}
\centering
\(
\scriptsize
    \begin{tabular}{p{9.7em}|p{3.2em}<{\centering}p{4.2em}<{\centering}|b{3.2em}<{\centering}p{4.2em}<{\centering}|b{3.2em}<{\centering}p{4.2em}<{\centering}}
    	\thickhline   & \multicolumn{2}{c|}{\textbf{Test1(head)}} & \multicolumn{2}{c|}{\textbf{Test2(tail)}} & \multicolumn{2}{c}{\textbf{Test3(random)}} \\
    	              & NDCG@5         & NDCG@10                  & NDCG@5         & NDCG@10                  & NDCG@5         & NDCG@10                   \\\hline
    	BM25          & \textbf{54.27} & \textbf{56.62}           & \textbf{38.42} & \textbf{41.82}           & \textbf{44.66} & \textbf{47.07}            \\
    	BM25-NP@WQS-P & 53.53          & 55.92                    & 37.39          & 40.78                    & 43.86          & 46.41                     \\
    	BM25-NP@BWC-P & 52.57          & 55.05                    & 35.68          & 39.26                    & 42.71          & 45.27                     \\
    	BM25-WT-P     & 53.62          & 56.02                    & 36.15          & 39.52                    & 42.95          & 45.43                     \\
    	BM25-WBN-P    & 53.62          & 56.29                    & 37.23          & 40.44                    & 42.96          & 45.70                     \\
    	BM25-RR@WQS-P & 53.78          & 56.29                    & 38.00          & 41.60                    & 43.88          & 46.31                     \\
    	BM25-RR@BWC-P & 54.23          & \textbf{56.62}           & 38.09          & 41.62                    & 44.33          & 46.65                     \\ \hline
    	KN            & \textbf{55.78} & \textbf{58.38}           & \textbf{40.23} & \textbf{43.90}           & \textbf{46.87} & \textbf{49.20}            \\
    	KN-NP@WQS-P   & 53.38          & 56.22                    & 39.05          & 41.92                    & 42.35          & 46.27                     \\
    	KN-NP@BWC-P   & 47.74          & 50.79                    & 35.95          & 40.12                    & 40.77          & 43.72                     \\
    	KN-WT-P       & 49.64          & 53.15                    & 38.22          & 42.16                    & 41.55          & 44.69                     \\
    	KN-WBN-P      & 48.86          & 52.40                    & 36.33          & 40.29                    & 40.65          & 43.78                     \\
    	KN-RR@WQS-P   & 48.91          & 52.42                    & 37.59          & 41.45                    & 41.64          & 44.58                     \\
    	KN-RR@BWC-P   & 51.97          & 54.83                    & 38.84          & 42.71                    & 41.70          & 44.78                     \\ \hline
    	DM            & \textbf{57.71} & 60.16                    & \textbf{37.82} & \textbf{41.25}           & \textbf{47.80} & \textbf{50.07}            \\
    	DM-NP@WQS-P   & 56.97          & 59.47                    & 35.78          & 39.64                    & 46.65          & 48.57                     \\
    	DM-NP@BWC-P   & 55.80          & 58.15                    & 34.85          & 38.40                    & 46.44          & 48.15                     \\
    	DM-WT-P       & 56.05          & 58.40                    & 34.97          & 38.49                    & 46.74          & 48.55                     \\
    	DM-WBN-P      & 56.06          & 58.48                    & 36.34          & 39.80                    & 46.88          & 48.98                     \\
    	DM-RR@WQS-P   & 57.39          & \textbf{60.26}           & 37.53          & 40.95                    & 46.95          & 49.08                     \\
    	DM-RR@BWC-P   & 57.04          & 60.04                    & 37.74          & 41.10                    & 46.90          & 49.03                     \\
    	\thickhline
    \end{tabular}
\)\\
\textbf{Bold}: the maximum value of the scheme for the dataset.
\end{center}
\end{table}

We find that most of the alternative methods perform worse and even statistically significantly worse than the baselines, except several measures. Especially on Test2, all of the alternative methods are worse than the baseline methods. All the alternative methods perform statistically significantly worse than the methods which utilize both query words and query phrases.

We find that there are several reasons.

(1) It appears that simply replacing query words with query phrases after query segmentation is not always helpful, and sometimes it is even harmful. For example, there is an NDCG loss for query ``cambridge university students count". The query segmentation result is ``cambridge university / students / count". When ``cambridge university" is combined together, it will not match with ``Cambridge" in a webpage, which also means Cambridge University.

(2) Incorrect segmentation is inevitable. Incorrect segmentation includes incorrect splitting of phrases such as ``my heart / will / go on / mp3 / download" and incorrect merging of words such as ``beijing seven eleven / stores". Both results can increase the number of incorrect phrase n-grams and reduce the number of correct phrase n-grams.

(3) Test2 consists of tail queries and it is difficult to conduct segmentations on such queries, because less information is available for the queries. Therefore, solely using phrases would not generate good performance in such case.

The experiment results demonstrate that it is better to make use of both query words and query phrases when employing query segmentation in relevance ranking.

\subsection{Comparison among Segmentation Methods}\label{sec:compare_with_diff_seg}

The results in Table~\ref{tab:rr_seg} also show differences in performances by employing different query segmentation methods in our approach of relevance ranking on Test1, Test2 and Test3.

The six segmentation methods (i.e. NP, WT, WBN and RR on BWC and WQS) have different relevance results. Among the segmentation methods, our method of re-ranking trained with the BWC dataset (*-RR@BWC-WP) achieves the best performances on nearly all three ranking schemes of three test data sets. One exception is WT on Test2 Data which has the best performance.

We conduct analysis on how segmentation accuracy and average segment length affect the performance of relevance ranking.

\subsubsection{Impact of Segmentation Accuracy}\label{sec:accuracyinfluence}
\begin{figure}[!t]
\centering
\includegraphics[width=120mm]{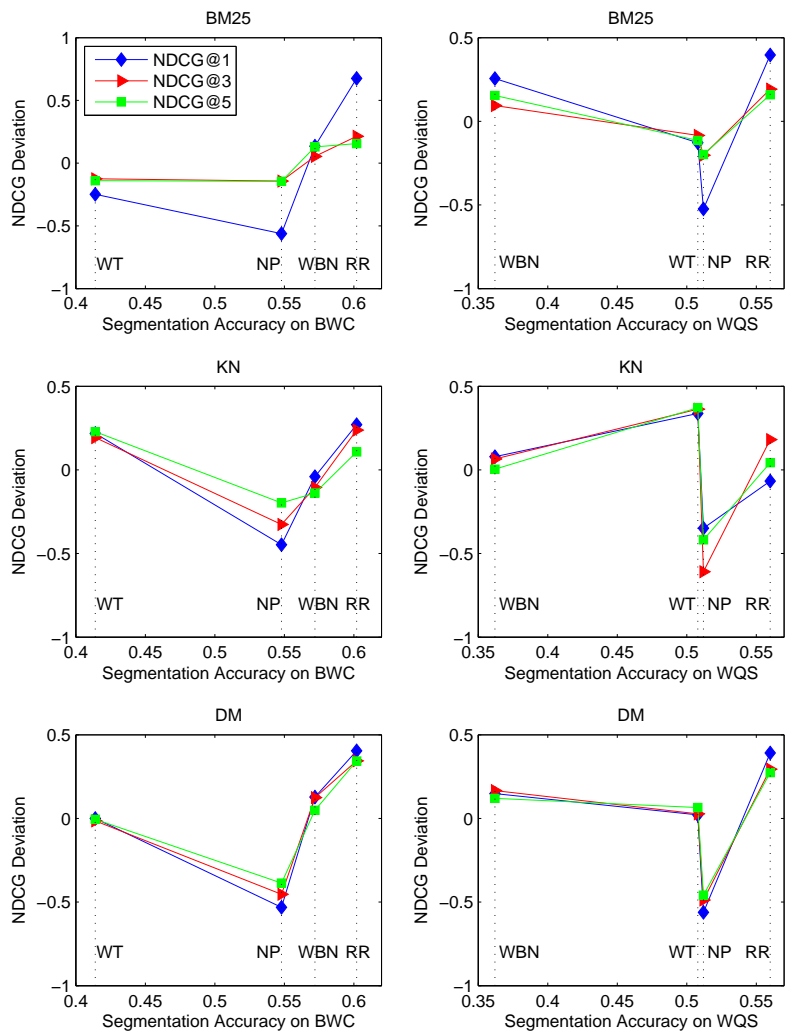}
\caption{\label{fig:QueryAccNdcgDeviation}The impact of query segmentation accuracy on relevance ranking performance.}
\end{figure}

We investigate whether higher query segmentation accuracy can generate higher relevance ranking performance. Figure~\ref{fig:QueryAccNdcgDeviation} shows different relevance ranking results of four basic segmentation methods. Column 1 shows the impact of segmentation accuracy on the BWC data, column 2 shows the impact of segmentation accuracy on the WQS data. We simply use query segmentation accuracy as measure because other measures have similar trends. Note that in Figure~\ref{fig:QueryAccNdcgDeviation}, the vertical axis represents the deviations of NDCG scores, which is defined as NDCG@i-Avg(NDCG@i). We use this measure to highlight the gradual change of NDCG with respect to query segmentation accuracy.

We can see that our re-ranking segmentation methods (RR@BWC and RR@WQS), which have the most accurate segmentation accuracies, achieve the highest relevance ranking performances. On the other hand, the relevance ranking performances do not always increase when more accurate segmentation methods are employed, which is consistent to the observations in the previous work of \citet{Hagen2012,SahaRoy2012}. For example, NP is a more accurate segmentation method than WT on WQS, however, the relevance ranking performances of NP@WQS are worse than those of WT.

There are three reasons for the inconsistency between segmentation accuracy and relevance ranking performance.

First, the training data of BWC and WQS are sampled from search log under several constraints~\citep{Hagen2011}, for example ``the query should consist of only determiners, adjectives, and nouns". These constraints make the queries in BWC and WQS different from the queries in relevance ranking, which are randomly sampled without any constraints. Therefore more accurate segmentation on BWC and WQS cannot guarantee better relevance ranking results .

Second, segmentation methods may suffer from over fitting, which indicates that a high segmentation accuracy does not necessarily mean a high relevance ranking accuracy. For example, NP makes use of many carefully designed features, such as whether the token is ``the" or ``free", because the authors find that ``the" and ``free" in the training data are likely to be split into single segments. Although these features are helpful for query segmentation, some of them may not help improve relevance.

Third, in some special cases, segmentation accuracy is not the most important factor for relevance ranking. For example, from Table~\ref{tab:rr_seg} we can see that WT has the best performance for KN on Test2 which contains only tail queries, although WT is not the best segmentation method on both BWC and WQS. We look into the extracted key n-grams of documents and find that 39.6\% n-grams are Wikipedia titles. The key n-gram model makes use of extracted unigrams, bigrams, and trigrams. WT tends to create segmentations consisting of Wikipedia titles and one-word segments. As a result, it is likely that the query representations of WT and the key n-gram fields match well in relevance ranking.

\subsubsection{Impact of Average Segment Length}
\begin{figure}[!t]
\includegraphics[width=110mm]{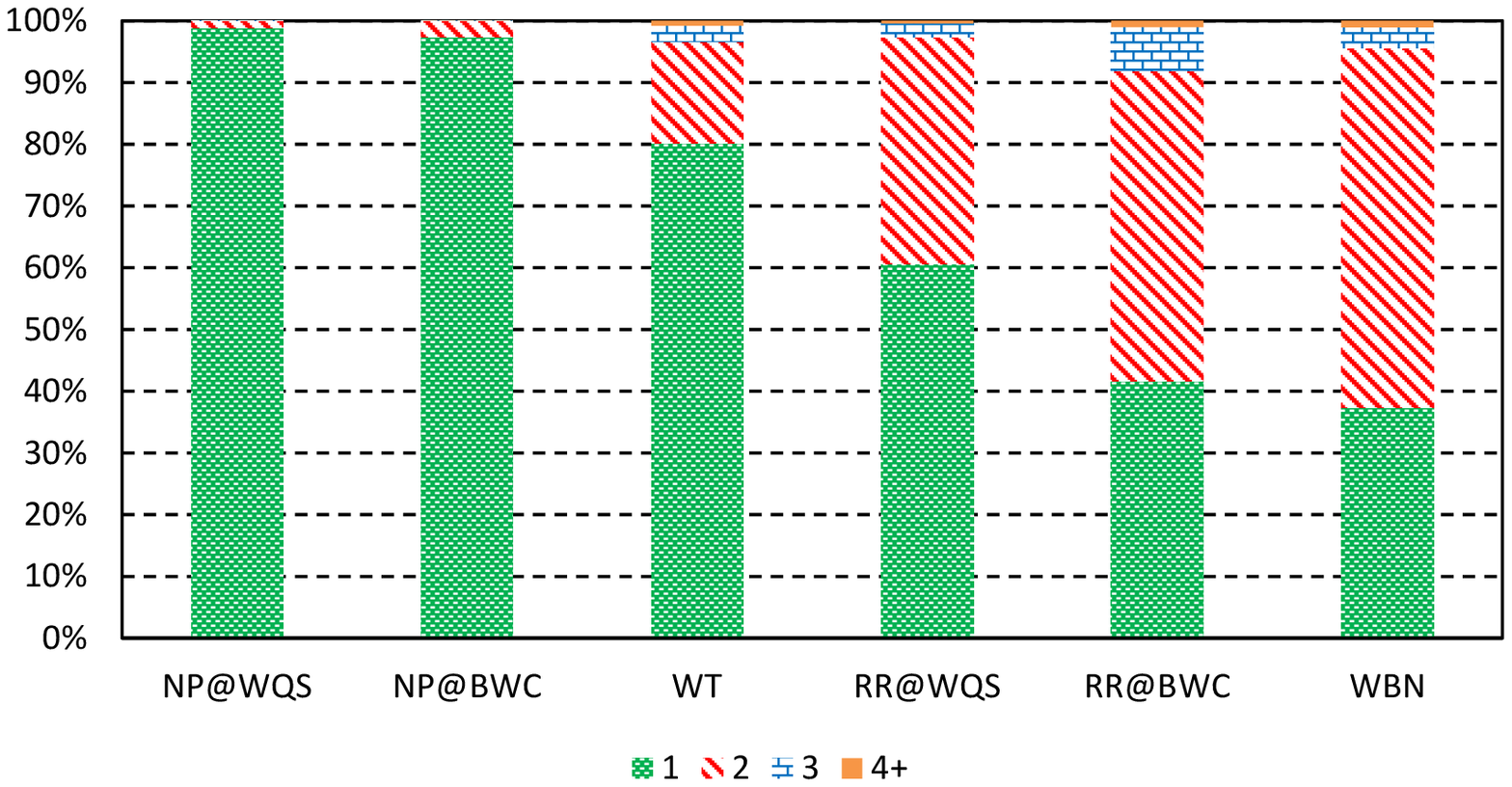}
\caption{\label{fig:SegmentLengthDistribute}Distributions of segment lengths by different segmentation methods.}		
\end{figure}

We investigate the impact of segment lengths on the performances of relevance ranking. Figure~\ref{fig:SegmentLengthDistribute} shows the distributions of segment lengths by the four methods. NP@BWC, NP@WQS and WT tend to create short segments, while RR@BWC, RR@WQS and WBN tend to create long segments. There are two reasons for the different segment distributions: different characteristics of segmentation methods and different training sets. First, it is likely for NP to generate shorter segments because it only considers the frequencies of two adjacent words~\citep{Bergsma07}, without considering whether the two words is a part of Wikipedia title. These methods tend to break entity names with low frequencies into single words, such as ``xbox one", ``final fantasy 7"  For WT, it is likely to treat Wikipedia titles as segments and make the rest one-word segments, yielding fine-grained segmentations. In contrast, WBN and our re-ranking method are likely to merge Wikipedia titles as well as the consecutive words with high frequencies as segments, yielding coarse-grained segmentations. Second, the two training datasets, BWC and WQS, have different average segment lengths (See Table~\ref{tab:segdist}), BWC has longer segments, while WQS has shorter segments. Therefore, the lengths of segments created by the supervised segmentation methods of NP and RR are quite different. As a result, NP@BWC and RR@BWC generate longer segments than NP@WQS and RR@WQS respectively.

\begin{figure}[!t]
\includegraphics[width=120mm]{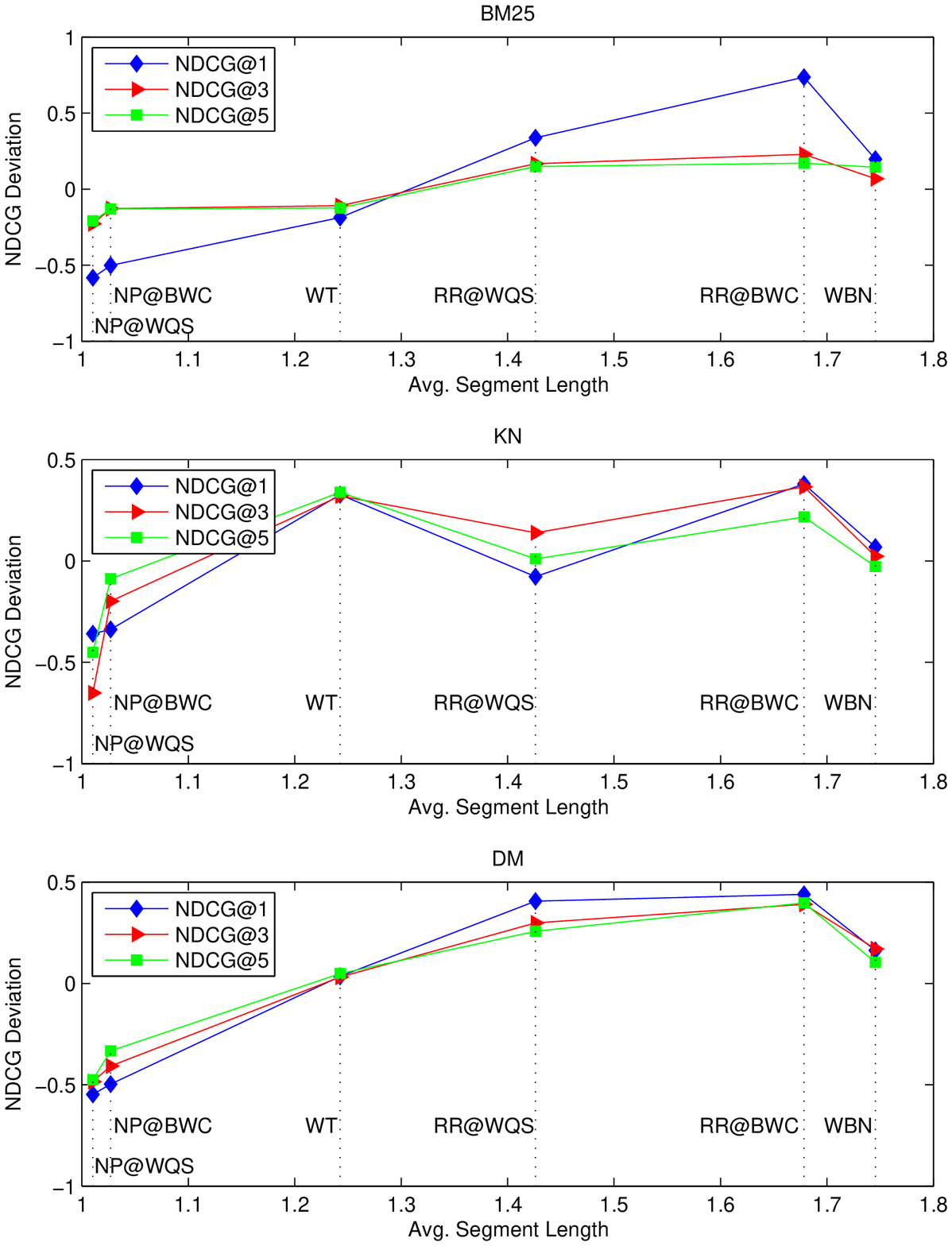}	
\caption{\label{fig:AvgLengthNdcgDeviation}The impact of average segment length on relevance ranking.}
\end{figure}

We observe that there is a tendency that coarse-grained segmentation outperforms fine-grained segmentation.  It is easy to understand that using a fully segmented query is equivalent to using all the query words. Therefore, the performance of using the former in relevance ranking will be the same as that of using the latter. That is why NP@WQS and NP@BWC do not have much space to improve. In contrast, the coarse-grained segmentation methods, which combine those highly associated words into a segment, can add useful information for matching between query and document, and thus generate higher relevance ranking accuracy. However, if the segmentation method gives too coarse segmentations, such as WBN, it also hinders the improvements. Figure~\ref{fig:AvgLengthNdcgDeviation} shows the relations between the deviations of NDCG scores with respect to average segment lengths, in the same manner as in Section~\ref{sec:accuracyinfluence}.

\section{Conclusions}\label{sec:conclusions}
In this paper, we have proposed a new approach to enhancing search relevance by query segmentation, including a method of query segmentation and a method of using query segmentation result in search. The former method first generates top $k$ candidates for query segmentation with a generative model and then re-ranks the candidates with a discriminative model. The latter method takes both the original query words and the segmented query phrases as units of query representation. We have empirically studied the effectiveness of the proposed approach with the relevance schemes of BM25, key n-gram model, and dependency model, and with one large scale dataset and two benchmark datasets. We have found that our approach can statistically significantly improve relevance ranking.

\bibliographystyle{spbasic}      
\bibliography{ref_short_short}

\end{document}